\documentclass[aps,prb, twocolumn, superscriptaddress, amsmath, amsfonts, amssymb, amsthm]{revtex4-2}

\usepackage{graphicx}
\usepackage{dcolumn}
\usepackage[unicode=true,
 bookmarks=false,
 breaklinks=true,pdfborder={0 0 1},backref=false,colorlinks=true]
 {hyperref}
\hypersetup{pdfcreator={},pdfproducer={LaTeX with hyperref},linkcolor=blue,anchorcolor=blue,citecolor=blue,filecolor=red,menucolor=red,urlcolor=blue,pdfstartview=FitV,pdfhighlight=/I,hypertexnames=true}

\usepackage{lmodern}
\usepackage{mathptmx, newtxtext, newtxmath, xspace}
\usepackage{colortbl}
\usepackage{array}
\usepackage{bbm}
\usepackage{siunitx}
\usepackage{comment}
\usepackage[table, dvipsnames]{xcolor}

\usepackage{bm}
\usepackage{braket}

\usepackage{ragged2e}

\usepackage{natbib}
\setcitestyle{numbers}

\renewcommand{\eqref}[1]{Eq. (\ref{#1})}

\begin{document}
\title{Compatible Instability: Gauge Constraints of Elasticity Inherited by Electronic Nematic Criticality}
\author{W. Joe Meese}
\affiliation{Department of Physics, The Grainger College of Engineering, University of Illinois Urbana-Champaign, Urbana, IL 61801, USA}
\affiliation{Anthony J. Leggett Institute for Condensed Matter Theory, The Grainger College of Engineering, University of Illinois Urbana-Champaign, Urbana, IL 61801, USA}
\author{Rafael M. Fernandes}
\affiliation{Department of Physics, The Grainger College of Engineering, University of Illinois Urbana-Champaign, Urbana, IL 61801, USA}
\affiliation{Anthony J. Leggett Institute for Condensed Matter Theory, The Grainger College of Engineering, University of Illinois Urbana-Champaign, Urbana, IL 61801, USA}
\date{March 10, 2026}
\begin{abstract}
Electronic nematicity is widely observed in quantum materials with varying degrees of electronic correlation, manifesting through charge, spin, orbital, or superconducting degrees of freedom. A phenomenological model capable of describing this broad set of systems must also account for nemato-elasticity, by which nematic and elastic degrees of freedom become intertwined. However, being a tensor gauge field theory, elasticity must satisfy the compatibility relations which guarantee the integrability of lattice deformations. Here, we develop a formalism for nemato-elasticity that manifestly respects the elastic compatibility relations. We show that these constraints bifurcate the phase space of nematic fluctuations into two orthogonal sectors: one compatible and thus critical, the other incompatible and therefore gapped. The suppression of the latter leads to universal direction-selective nematic criticality in any crystal lattice. Moreover, the critical nematic modes are protected from pinning effects induced by microscopic defect strains, which necessarily induce both longitudinal and transverse correlated random fields. Finally, our results also reconcile seemingly contradictory nematic phenomena, such as the mean-field character of the nematic transition and the widespread presence of domain formation. 
\end{abstract}
\maketitle

Electronic nematicity, the spontaneous breaking of rotational symmetry by electronic degrees of freedom, has been reported in a wide range of systems such as 2DEGs \cite{Eisenstein99_nem,Feldman16_nem},
iron-based superconductors \cite{Chu2010_nem,Chuang2010_nem,Fernandes2014},
cuprates \cite{Hinkov08_nem,Davis10_nem}, correlated $f$-electron
materials \cite{Ronning17_nem,Rosenberg2019_nem,Seo2020_nem,Massat2022_nem,Jiang2025},
doped topological insulators \cite{Tamegai2019_nem,Cho2020_nem},
flat-band kagome metals \cite{Drucker2024_nem}, 2D van der Waals
antiferromagnets \cite{RMF08_Potts_nematic_strain,Hwangbo2024_nem,Sun2024_nem,Tan2024_nem},
Bernal bilayer graphene \cite{Geim11_nem}, transition metal dichalcogenides
\cite{Little2020_nem,Hamill2021_nem,Cho2022_nem,Silber2024_nem},
colossal magnetoresistance materials \cite{Beaudin2022_nem}, ruthenates
\cite{Mackenzie07_nem}, twisted moir\'e systems \cite{Jiang2019_nem,Cao2021_nem,RubioVerdu2022_nem,Zhang2022_nem},
and topological semimetals \cite{Siddiquee2022_nem,Venkatesan2024_nem,Hossain2024_nem}. Since this vast list encompasses materials with very distinct properties (metals, insulators, and superconductors), a phenomenological description of electronic nematicity is necessary to gain a broad understanding of this phenomenon. In analogy to the phenomenological theory of classical liquid crystals, one can define a symmetric and traceless rank-2 tensorial order parameter $\varphi_{ij}$ \cite{Oganesyan01}, whose five independent components correspond to the many-body expectation values of quadrupolar order in the charge, orbital, magnetic, or superconducting sector.  They are usually expressed as in-plane $\varphi_{x^2-y^2}, \varphi_{xy}$ and out-of-plane  $\varphi_{z^2}$, $\varphi_{yz}$, $\varphi_{xz}$ quadrupolar order parameters, where the subscripts refer to the standard $d$-wave form factors in Cartesian coordinates.

A crucial difference with respect to liquid crystals is that electronic nematicity takes place on a lattice, which explicitly breaks rotational symmetry. Phenomenologically, this effect is captured through a bilinear coupling between  $\varphi_{ij}$  and the traceless part of the elastic tensor $\varepsilon_{ij}-\frac{1}{3}\varepsilon_{ii} \delta_{ij}$. The latter is defined as $\varepsilon_{ij}=\tfrac{1}{2}(\partial_i u_j + \partial_j u_i)$, where the lattice displacement vector $ \boldsymbol{u}(\boldsymbol{x})$ describes the geometric deformation of the medium by displacing every element from $\boldsymbol{x}$ to $\boldsymbol{x} + \boldsymbol{u}(\boldsymbol{x})$. Some consequences of this coupling between nematicity and physical displacements of the lattice, dubbed nemato-elastic coupling, are well-known in the literature.  For instance,  nematic order triggers a symmetry-breaking lattice distortion that lowers the point group of the crystal \cite{Fradkin_review,Fernandes2014}. Conversely, the rotational symmetry-breaking promoted by nematic order can only be discrete, and the continuous nematic order parameter is split in $q$-state Potts or clock order parameters \cite{RMF14_classification_nematics}. 

Through this nemato-elastic coupling, two different types of lattice fluctuations impact the electronic nematic critical properties in sharply distinct ways. Elastic \textit{thermal fluctuations} generate effective long-ranged nematic interactions, resulting in a sharp, mean-field like nematic transition \cite{Karahasanovic16, Paul17, Fernandes2020, sanchezQuantitativeRelationshipStructural2022, Hecker2022, stewardElasticQuantumCriticality2025}. Conversely, there are also \textit{statistical fluctuations} associated with the random strain fields generated by quenched crystalline defects \citep{tanatarDirectImagingStructural2009, jescheCouplingStructuralMagnetic2010, Ran2011, Forrest2016, Ren2021, bohmerNematicityNematicFluctuations2022, curroNematicityGlassyBehavior2022}. These random strains locally break the rotational symmetry of the lattice and couple to the nematic order parameter as a random conjugate field, an effect known to destabilize long-range order in favor of domain formation \cite{schneiderRandomfieldInstabilityFerromagnetic1977, Imry_Ma, Binder1983, nattermannInstabilitiesIsingSystems1983, grinsteinSurfaceTensionRoughening1983, Carlson2006, Dahmen2010, carlsonUsingDisorderDetect2011, Phillabaum2012, vojtaPhasesPhaseTransitions2013, nieQuenchedDisorderVestigial2014, vojtaDisorderQuantumManyBody2019, mirandaPhaseDiagramFrustrated2021, yeStripeOrderImpurities2022, RMF01_RFBM, RMF06_Potts_nematic_strain,yangCoarseningDynamicsIsingnematic2025}.
This leads to a seemingly contradictory scenario in which the \textit{same} nemato-elastic coupling favors \textit{both} mean-field nematic behavior (and hence a sharp transition) \textit{and} nematic domain breakup (and hence a smeared transition). This contradiction is further exacerbated by experiments, as macroscopic probes observe typical signatures of a sharp mean-field nematic transition \citep{Ran2011, Forrest2016, Chu2010_nem, Chu12_nem, bohmerNematicityNematicFluctuations2022, Kuo16_nem, Hwangbo2024_nem}, whereas local probes often report mesoscopic nematic domain formation due to disorder \citep{tanatarDirectImagingStructural2009, Ran2011, Forrest2016, Ren2021, Dioguardi2015, Dioguardi2016, Hwangbo2024_nem, maMicrostructureTetragonaltoorthorhombicPhase2009, Rosenthal2014, RMF08_Potts_nematic_strain}. Addressing this fundamental problem in our understanding of electronic nematic phenomena requires a theoretical approach that treats the intertwined nematic and elastic degrees of freedom on an equal footing.

In this paper, we develop such an approach for electronic nematic phase transitions that take place at nonzero temperatures. While quantum critical nematicity is of obvious interest \citep{chandraIsingTransitionFrustrated1990, readLargeNExpansionFrustrated1991, gorkovNontrivialMagneticOrder1992, kivelsonElectronicLiquidcrystalPhases1998, fradkinLiquidcrystalPhasesQuantum1999, halboth$d$WaveSuperconductivityPomeranchuk2000, fradkinNematicPhaseTwoDimensional2000, yamaseInstabilityFormationQuasiOneDimensional2000, zaanenDuality2+2004, dellannaFermiSurfaceFluctuations2006, wuFermiLiquidInstabilities2007, xuIsingSpinOrders2008, qiGlobalPhaseDiagram2009, zachariasMultiscaleQuantumCriticality2009, maslovFermiLiquidPomeranchuk2010, garstElectronSelfenergyNematic2010, metlitskiQuantumPhaseTransitions2010, yamaseNematicQuantumCriticality2011, kieselUnconventionalFermiSurface2013, maciejkoFieldTheoryQuantum2013, youTheoryNematicFractional2014, metlitskiCooperPairingNonFermi2015, yamaseElectronicNematicPhase2015, boettcherUnconventionalSuperconductivityLuttinger2018, yamaseTheoreticalInsightsElectronic2021}, the role of dynamics lies outside the scope of the present work. Instead, we focus on the classical field theoretical description and derive a consistent model for nemato-elastic phenomena by embracing a key aspect of elasticity: being a theory of geometric deformations, elasticity is a tensor gauge field theory.
Much like gauge invariance in electrostatics imposes constraints on the electric field encoded in $(\boldsymbol{\nabla} \times \boldsymbol{E})_i = \epsilon_{ijk}\partial_j E_k = 0$, where $\epsilon_{ijk}$ is the Levi-Civita symbol, gauge invariance in elasticity enforces constraints on the strain tensor that can be written concisely as $\text{inc}(\varepsilon)_{ij} \equiv \epsilon_{ikl}\epsilon_{jmn}\partial_k\partial_m\varepsilon_{ln} = 0$ \citep{kronerekkehartContinuumTheoryDefects1981, Dewit1973, kleinertGaugeFieldsSolids1989}. Here, $\text{inc}(\varepsilon)_{ij}$ is known as the incompatibility operator, and the three independent second-order partial differential equations that follow from vanishing incompatible strain are known as the Saint Venant compatibility relations \cite{kronerekkehartContinuumTheoryDefects1981,muratoshioMicromechanicsDefectsSolids1987}. Ultimately, these constraints are required because  $\boldsymbol{u}$  has \textit{three} independent degrees of freedom while $\varepsilon_{ij}$ has \textit{six}. 

The compatibility relations are known to have a pronounced effect on various types of phase transition, such as the martensitic
\cite{Kartha1995,Shenoy1999,Rasmussen2001,Lookman2003}, the metal-to-insulator
\cite{Guzman2019}, and the ferroelectric \cite{Littlewood2014,Kimber2023}. Moreover, the strain field generated by lattice defects such as dislocations, disclinations, and vacancies fail to satisfy the compatibility relations, and are thus described in terms of a non-zero incompatibility tensor \citep{eshelbyContinuumTheoryLattice1956, dewitLinearTheoryStatic1970, Dewit1973, dewitTheoryDisclinationsIII1973, dewitTheoryDisclinationsIV1973, kronerekkehartContinuumTheoryDefects1981, kleinertGaugeFieldsSolids1989, valsakumarGaugeTheoryDefects1988, beekmanDualGaugeField2017, Pretko2018_PRL, Pretko2019, gaaFractonelasticityDualityTwisted2021}. Nevertheless, their impact on electronic nematicity remains little explored. While it has been recognized that residual strain generated by these unavoidable crystalline defects act as a random conjugate field to the nematic order parameter, and that they can significantly impact the nematic critical properties \cite{Carlson2006,Phillabaum2012,RMF01_RFBM}, the inhomogeneous strains are often assumed to be random and uncorrelated, which is inconsistent with the properties of the incompatibility tensor.

To illustrate how the compatibility relations fundamentally alter nematic critical properties, consider for example the condition encoded in  $\text{inc}(\varepsilon)_{33} = 0$: 
\begin{equation}
\nabla^{2}\left(\varepsilon_{xx}+\varepsilon_{yy}\right)=\left(\partial_{x}^{2}-\partial_{y}^{2}\right)\left(\varepsilon_{xx}-\varepsilon_{yy}\right)+\left(2\partial_{x}\partial_{y}\right)\left(2\varepsilon_{xy}\right).\label{eq:2D_SVCR}
\end{equation}
This relationship establishes the interdependence between the symmetry-preserving, volume-changing \textit{dilatation} strain component $\varepsilon_{xx}+\varepsilon_{yy}$ and the in-plane symmetry-breaking, volume-preserving \textit{deviatoric}  $\varepsilon_{xx} - \varepsilon_{yy}$ and \textit{shear} $\varepsilon_{xy}$ strain components. Despite the fact that these different types of strain have different symmetry properties (i.e., transform as different irreducible representations in group-theory language), they become entangled by the compatibility relations for all \textit{inhomogeneous} deformations. Importantly, while the deviatoric and shear strains have a symmetry-allowed coupling to the nematic order parameter, the dilatation does not.  However, the compatibility relation in \eqref{eq:2D_SVCR} shows that generic spatial fluctuations of the nematic order parameter will generate dilatations, meaning that nematic fluctuations can incur a higher elastic energy cost than expected along symmetry grounds alone.

\begin{figure}
    \centering
    \includegraphics[width=\columnwidth]{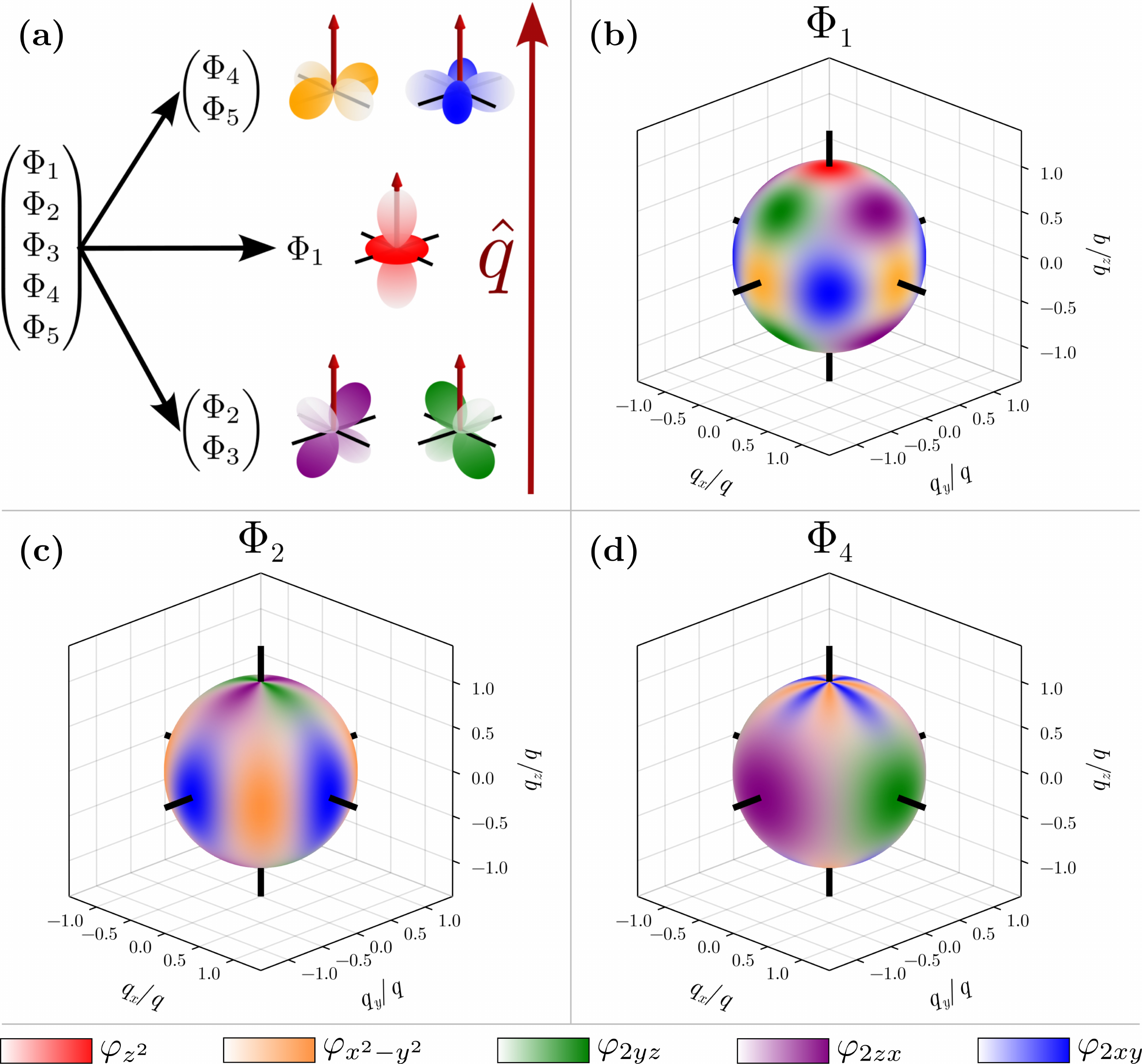}
    \caption{\justifying Nonlocal transformations between the $d$-orbital nematic basis, $\boldsymbol{\varphi}$, and the co-rotating helical basis, $\boldsymbol{\Phi}$.
    Along each $\hat{q}$, the fivefold degenerate electronic nematic order parameter in the helical basis is split by the compatibility relations. \textbf{(a)} The splitting is shown for $\hat{q} = \hat{z}$. In this case, the $d$-orbital doublets $(\varphi_{2yz}, \varphi_{2zx})$ and $(\varphi_{x^2-y^2}, \varphi_{2xy})$  map onto $(\Phi_{2}, \Phi_{3})$ and $ (\Phi_{4}, \Phi_{5})$, respectively, while  $\varphi_{z^2}$ maps onto $\Phi_1$. \textbf{(b, c, d)} Mappings from the nematic $d$-orbital order parameter $\boldsymbol{\varphi}$ onto selected nematic helical order parameters, $\Phi_{1,2,4}$, as functions of momentum. The colormaps indicate when each $\Phi_{1,2,4}$  function has the strongest overlap with each of the $d$-orbital basis functions from $\boldsymbol{\varphi}$. 
    }
    \label{fig:PhiAmplitude_Decomp_Orbitals}
\end{figure}

In this Letter and in the accompanying paper \citep{LongPaper}, we develop a general phenomenological theory of nemato-elasticity that automatically obeys the non-analytic geometrical constraints enforced by the compatibility relations for electronic nematic fluctuations with infinitesimal, but nonzero, momenta.  This is accomplished by expressing the five components of the nematic order parameter $\varphi_{ij}$  not in the usual quadrupolar Cartesian basis $\{\varphi_{x^2-y^2}, \varphi_{xy},\varphi_{z^2}, \varphi_{yz}, \varphi_{xz} \}$ (which we dub ``$d$-orbital basis''), but in a new ``helical basis'' $\{\Phi_a\}$  (with $a=1,...,5$) \citep{RMF14_classification_nematics, LongPaper}. Its defining property is that the five order parameters are orthogonal for every momentum $\boldsymbol{q}$. As a result, the transformation between bases is nonlocal in real-space, and the $d$-orbital content of a given helical nematic order parameter $\Phi_a$ changes depending on its momentum direction $\hat{q}$, as sketched in Fig. \ref{fig:PhiAmplitude_Decomp_Orbitals}. The main advantage of using the helical basis is that one can immediately separate the strain components that satisfy the compatibility conditions from those that do not. In contrast, in the $d$-orbital basis, one would need to ensure via complicated boundary conditions that the compatibility conditions are satisfied.

Employing this helical basis for the general case of a three-dimensional isotropic solid undergoing a nematic transition, we demonstrate four main consequences of the compatibility relations on nemato-elastic criticality. (i) Only two of the five helical order parameters are ever critical, and they map onto a single orbital nematic order parameter only along particular momentum directions. (ii) Direction-selective criticality is a universal property of the nematic transition regardless of the underlying crystal structure. While previous works have attributed this phenomenon to the anisotropy of the phonon dispersion in certain crystals \cite{Karahasanovic16,Paul17,Fernandes2020}, here we show that it actually originates from the prohibition of incompatible nematic fluctuations. (iii) Structural disorder, quantified through the violation of the compatibility relations, couples only to the three non-critical helical nematic order parameters. These three components are thus responsible for the phenomenon of nemato-plasticity. (iv) Projection of the three incompatible nematic order parameters onto the orbital basis reveals domain breakup due to the emergence of long-range correlated longitudinal and transverse random fields associated with incompatible strain and crystalline defects. 

We demonstrate these results by considering nemato-elasticity in an isotropic elastic medium. The electronic nematic order parameter transforms as the $\ell = 2$ irreducible representation of the rotation group $\text{SO}(3)$, i.e., as a quadrupolar order parameter. It is thus defined, for example, by the traceless rank-2 symmetric tensor
\begin{equation}
    \varphi_{ij}\left(\boldsymbol{x}\right)=-\left\langle \hat{\psi}^{\dagger}\left(\boldsymbol{x}\right)\left(\partial_{i}\partial_{j}-\tfrac{1}{3}\delta_{ij}\nabla^{2}\right)\hat{\psi}\left(\boldsymbol{x}\right)\right\rangle , \label{eq:electronic_nematic_tensor}
\end{equation}
where $\hat{\psi}(\boldsymbol{x})$ is the electronic wavefunction in real-space \citep{Oganesyan01, Fradkin_review, RMF14_classification_nematics}. It is straightforward to generalize this expression for orbital, spin, or superconducting operators. This tensor can be reorganized as the five-component vector 
$\boldsymbol{\varphi}\equiv( \varphi_{z^2}, \varphi_{x^2-y^2}, \varphi_{2yz}, \varphi_{2zx}, \varphi_{2xy})^{\text{T}}$ whose components explicitly transform as the five electronic $d$-orbitals. In an anisotropic crystal, $\boldsymbol{\varphi}$ is split in irreducible representations of the point group \citep{RMF14_classification_nematics}. For instance, tetragonal systems like cuprates and iron-based superconductors have the familiar Ising-nematic order parameters $\varphi_{2xy} \equiv 2\varphi_{xy}$  or $\varphi_{x^2-y^2} \equiv \varphi_{xx}-\varphi_{yy}$ \citep{chubukovOriginNematicOrder2015, fernandesIronPnictidesChalcogenides2022,achkarNematicityStripeorderedCuprates2016}, whereas in hexagonal systems like Bi$_2$Se$_3$, these two planar nematic components transform as a doublet described in terms of a 3-state Potts order parameter  \citep{Cho2020_nem, RMF08_Potts_nematic_strain, Hwangbo2024_nem, RMF14_classification_nematics, galiCriticalNematicPhase2024, hattoriOrbitalMoireQuadrupolar2024}. To keep our considerations general and show that our results are not caused by the anisotropies of the lattice, but instead by the compatibility relations, we consider the ideal case of an isotropic medium. In this case, the strain tensor $\varepsilon_{ij}$ has six components, five of which comprise the symmetry-breaking, \textit{deviatoric} strains that also transform as the $\ell = 2$  irreducible representation of $\text{SO}(3)$, $\boldsymbol{\varepsilon}\equiv (\varepsilon_{z^2},\varepsilon_{x^2-y^2},\varepsilon_{2yz},\varepsilon_{2zx},\varepsilon_{2xy})^{\text{T}}$. Thus, the nemato-elastic coupling appears in the free energy as $f_{\text{ne}} = -\lambda_0 (\boldsymbol{\varphi}\cdot \boldsymbol{\varepsilon})$, where $\lambda_0$  is the single symmetry-allowed nemato-elastic coupling constant in $\text{SO}(3)$.

Given that the compatibility relations entangle symmetry-breaking deviatoric strain $\boldsymbol{\varepsilon}$ with symmetry-preserving dilatation strain $\varepsilon_0\equiv \varepsilon_{ii}$ for infinitesimal momentum (see Eq. \ref{eq:2D_SVCR}), they are cumbersome to be implemented in the $d$-orbital basis for nemato-elasticity. Instead, we define a basis that co-rotates with the momentum, which we dub the ``helical nematic" basis \citep{LongPaper}. It is constructed at each momentum $\boldsymbol{q}$ via an orthonormal set of basis vectors $\{\hat{e}_{1,2,3}\}$, with $\hat{e}_1 \equiv \boldsymbol{q}/q$ longitudinal to the momentum and $\hat{e}_2$, $\hat{e}_3$, corresponding to the transverse azimuthal and polar directions, respectively. In Fourier space, the three helical strain components $\varepsilon_\alpha^h \equiv \text{i}q(\hat{e}_\alpha \cdot \boldsymbol{u})$ are sufficient to completely determine the six strain components in the Cartesian basis  \citep{LongPaper}:
\begin{equation}
    \varepsilon_{ij}=\mathcal{Q}_{ij}^{1\alpha}\varepsilon_{\alpha}^{h},\quad\mathcal{Q}_{ij}^{\alpha\beta}\equiv\tfrac{1}{2}\left(\hat{e}_{\alpha,i}\hat{e}_{\beta,j}+\hat{e}_{\alpha,j}\hat{e}_{\beta,i}\right), \label{eq:def_helical_strain_Qabij}
\end{equation}
where we have omitted the $\boldsymbol{q}$-dependence for brevity and the sum over Greek indices is implicit ($\alpha=1,2,3$).  Crucially, the strain tensor only depends on \textit{three}  helical strain amplitudes, despite the fact it has \textit{six} components. By applying the incompatibility operator to $\mathcal{Q}^{\alpha\beta}$, one obtains $\text{inc}(\mathcal{Q}^{\alpha\beta})_{ij} = -q^2\epsilon_{1\alpha\gamma}\epsilon_{1\beta\delta}\mathcal{Q}^{\gamma\delta}_{ij}$. As a result, the compatibility relations are automatically satisfied in the helical basis.

\begin{figure*}
    \centering
    \includegraphics[width=\linewidth]{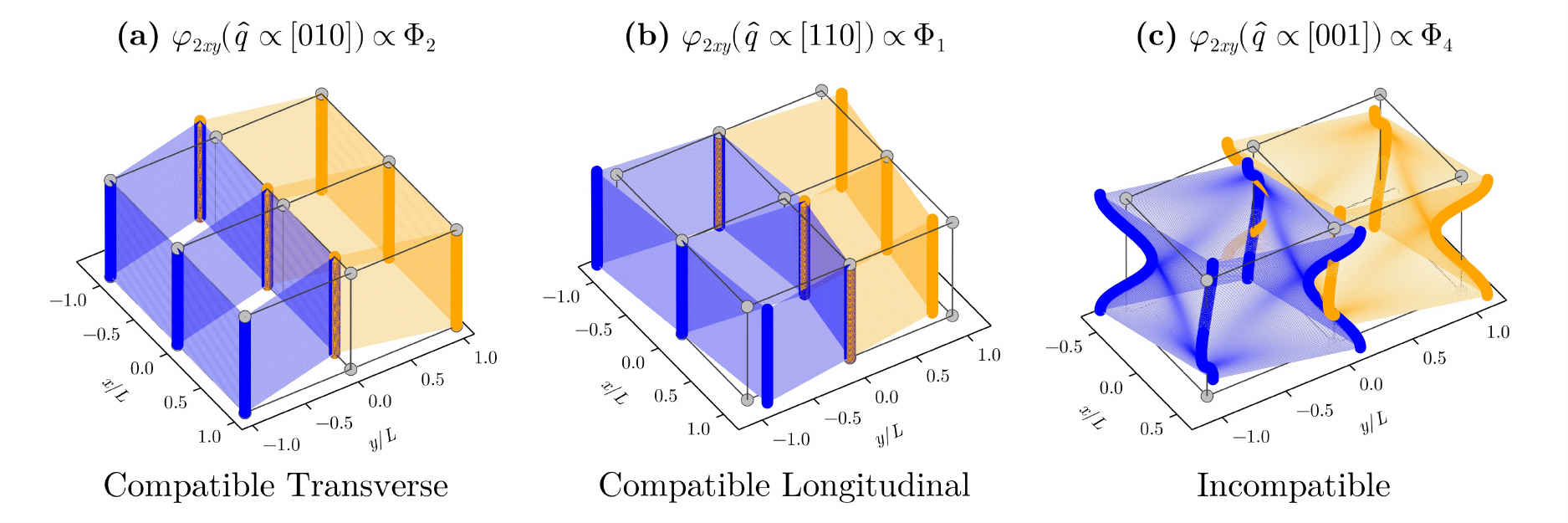}
    \caption{\justifying Schematic of compatible and incompatible nematic fluctuations. Each figure shows the distortion of a simple cubic grid (gray dots) of linear size $L$ in the presence of a nematic fluctuating mode $\varphi_{2xy}$ with wavevector along $\hat{q}$. 
    \textbf{(a)} Twin nematic domains are generated for $\boldsymbol{q} \propto [010]$, in which case $\varphi_{2xy} \propto \Phi_2$.
    \textbf{(b)} Energetically costly longitudinal elastic dilatation is generated for $\boldsymbol{q} \propto [110]$, in which case  $\varphi_{2xy}\propto \Phi_1$. \textbf{(c)} Incompatible deformations are generated when $\boldsymbol{q} \propto [001]$, for which $\varphi_{2xy}\propto \Phi_4$. The incompatibility manifests in the multi-valuedness required to displace the gray dots simultaneously to both the blue and orange positions. This leads to cracking and material overlap along the interface between blue and orange regions. }
    \label{fig:in-compatibility_side_by_side}
\end{figure*}

To make the connection between helical and $d$-orbital bases clearer, we use the set of five symmetric Gell-Mann matrices, $\boldsymbol{\lambda}\equiv\left(\lambda^{8},\lambda^{3},\lambda^{6},\lambda^{4},\lambda^{1}\right)^{\text{T}}$, to express the deviatoric strain tensor as $\boldsymbol{\varepsilon} = \boldsymbol{\lambda}_{ij}\varepsilon_{ij}$. In contrast, the longitudinal ($\varepsilon^h_1$)  and transverse ($\varepsilon^h_{2,3}$) strain amplitudes can be expressed in terms of the three orthonormal helical-basis vectors, $\boldsymbol{\hat{Q}}_1 = \frac{\sqrt{3}}{2} \boldsymbol{\lambda}_{ij}\mathcal{Q}_{ij}^{11}$, $\boldsymbol{\hat{Q}}_2 = \boldsymbol{\lambda}_{ij}\mathcal{Q}_{ij}^{12}$, and $\boldsymbol{\hat{Q}}_3 = \boldsymbol{\lambda}_{ij}\mathcal{Q}_{ij}^{13}$, with $\mathcal{Q}_{ij}^{\alpha\beta}$  defined in \eqref{eq:def_helical_strain_Qabij}.  These three vectors \textit{do not} form a complete basis in the five-dimensional quadrupolar space, however, even though they \textit{do} account for all compatible strains. The remaining two unit vectors are given by $\boldsymbol{\hat{Q}}_4 \equiv \boldsymbol{\lambda}_{ij}\mathcal{Q}_{ij}^{23}$  and $\boldsymbol{\hat{Q}}_5 = \frac{1}{2} \boldsymbol{\lambda}_{ij}(\mathcal{Q}_{ij}^{22} - \mathcal{Q}_{ij}^{33})$. These basis vectors are representations of the $\ell = 2$ basis functions (i.e., of any quadrupolar tensor) in the \textit{helical basis} at \textit{each} wavevector. When applied to nemato-elasticity, the nematic order parameter and deviatoric strain vectors are given by
\begin{equation}
    \boldsymbol{\varphi} = \sum_{a = 1}^5 \Phi_a \boldsymbol{\hat{Q}}_a, \quad \boldsymbol{\varepsilon} = \frac{2}{\sqrt{3}}\varepsilon^h_1 \boldsymbol{\hat{Q}}_1 + \sum_{a = 2}^3 \varepsilon^h_a \boldsymbol{\hat{Q}}_a.
\end{equation}
Hence, the nemato-elastic coupling in the free energy acquires the simplified form:
\begin{equation}
    \boldsymbol{\varphi}\cdot \boldsymbol{\varepsilon} = \tfrac{2}{\sqrt{3}}\Phi_{1}^{\phantom{h}}\varepsilon_{1}^{h}+\Phi_{2}^{\phantom{h}}\varepsilon_{2}^{h}+\Phi_{3}^{\phantom{h}}\varepsilon_{3}^{h}, \label{eq:iso_nematoelastic_bilinear}
\end{equation}
Unlike the compatible strain tensor, the electronic nematic order parameter fluctuates within the full five-dimensional $\ell = 2$ representation of $\text{SO}(3)$ at each $\boldsymbol{q}$,  being represented in the helical basis through the five amplitudes, $\{ \Phi_a\}$. The mapping between them and the $d$-orbital basis functions is shown in Fig. \ref{fig:PhiAmplitude_Decomp_Orbitals}. Panel (a) shows that, for a given momentum direction, the five helical components $\{ \Phi_a\}$ are split into one longitudinal ``singlet'' and two transverse ``doublets''. Panels (b)-(d) illustrate that, in general, a given nematic helical component $\Phi_a$ maps onto a combination of nematic $d$-orbital components $\varphi_b$, except along certain directions. For instance, $\Phi_1$ maps onto a pure $\varphi_{x^2-y^2}$ only along the $q_x=q_z=0$ and $q_y=q_z=0$ directions, and onto $\varphi_{2xy}$ along $q_x=\pm q_y$, $q_z=0$. 

Since the helical strain components satisfy the compatibility relations by construction, it is safe now to integrate them out and obtain the compatible nemato-elastic free energy. Considering an isotropic medium described by the Lam\'e coefficient,  $\lambda$, and the shear modulus,  $\mu$, we obtain a correction to the nematic free energy of the form
$\Delta f_{\text{nem}} = - \frac{1}{2} \boldsymbol{\varphi} \cdot \mathcal{M}(\hat{q}) \cdot \boldsymbol{\varphi}$. Here, $\mathcal{M}\left(\hat{q}\right)=\frac{\lambda_{0}^{2}}{\mu}(\boldsymbol{\hat{Q}}_{2}^{\phantom{\text{T}}}\boldsymbol{\hat{Q}}_{2}^{\text{T}}+\boldsymbol{\hat{Q}}_{3}^{\phantom{\text{T}}}\boldsymbol{\hat{Q}}_{3}^{\text{T}}+\varrho\boldsymbol{\hat{Q}}_{1}^{\phantom{\text{T}}}\boldsymbol{\hat{Q}}_{1}^{\text{T}})$ is a  $5\times 5$  matrix and $\varrho \equiv 4\mu/3(\lambda + 2\mu)\in (0,1)$  is a dimensionless, non-universal number. Using the completeness relation among the $\{\boldsymbol{\hat{Q}}_a\}$ basis vectors, we obtain for the effective nematic free energy: 
{\small
\begin{equation}
   f_{\text{eff}}=\tfrac{1}{2}\left\{ \left(r-\tfrac{\lambda_{0}^{2}}{\mu}+q^{2}\right)\boldsymbol{\varphi}\cdot\boldsymbol{\varphi}+\tfrac{\lambda_{0}^{2}}{\mu}\left[\left(1-\varrho\right)\Phi_{1}^{2}+\Phi_{4}^{2}+\Phi_{5}^{2}\right]\right\} +\dots.\label{eq:effective_nematic_free_energy_density}
\end{equation}}
\unskip\hspace{0pt}where $r$  measures the distance to the bare nematic critical point. The first term is an isotropic renormalization that enhances the bare nematic critical temperature from $r_c^0 = 0$ to the \textit{nemato-elastic} critical temperature,  $r_c = \lambda^2_0/\mu$.

More importantly, \eqref{eq:effective_nematic_free_energy_density} implies that the nematic criticality is direction-dependent. Indeed, the nematic instability occurs when the three helical nematic components $\Phi_1$, $\Phi_4$, and $\Phi_5$ simultaneously vanish, which can only happen along specific momentum-space directions, as illustrated in Fig. \ref{fig:PhiAmplitude_Decomp_Orbitals}. This also breaks the $\text{SO}(3)$  symmetry of the isotropic elastic medium, reducing the critical manifold to a two-dimensional space spanned by $\Phi_{2,3}$ with only $\text{SO}(2)$ symmetry. We dub this type of second-order phase transition \textit{compatible electronic nematic criticality}, which is a fundamental distinction between the phenomenologies of electronic nematicity in solids and nematicity in classical liquid crystals. Importantly, while a higher-order cubic invariant is allowed in the bare electronic nematic free energy on the isotropic lattice, we find that for a sufficiently small cubic coefficient, the interplay between the gapped and critical helical electronic nematic modes ensure that the transition remains second-order \cite{LongPaper}.

It is instructive to interpret these results in terms of the conventional orbital basis of nematicity, where each component $\varphi_a$ can be readily identified with a type of quadrupolar order. The implication  of \eqref{eq:effective_nematic_free_energy_density} is that the nematic instability described by a given  $\varphi_a$ can only be critical  when $\varphi_a$  completely overlaps with either $\Phi_2$  or $\Phi_3$. For example, the Ising-nematic order parameter $\varphi_{2xy}$ for tetragonal systems is only critical when the momentum lies along the $[100]$ or $[010]$  axes, in which case  $\varphi_{2xy}$ completely overlaps with $\Phi_2$, see  Fig. \ref{fig:PhiAmplitude_Decomp_Orbitals}(c) . In contrast, for momenta along the $[110]$ or $[\bar{1}10]$  axes,   $\varphi_{2xy}$ completely overlaps with $\Phi_1$, which incurs an additional energy cost according to  \eqref{eq:effective_nematic_free_energy_density}. Physically, this energy cost arises because fluctuating nematicity with these momenta directions generate, via the nemato-elastic coupling,  strain dilatation $\varepsilon^h_1$, in agreement with the compatibility relation, \eqref{eq:2D_SVCR}.  This is illustrated in  Fig. \ref{fig:in-compatibility_side_by_side}(b), which shows the strain pattern generated by the nematic mode $\varphi_{2xy}$ with a small wavevector along  $[110]$. In contrast, when the wavevector is along  $[100]$, as shown in  Fig. \ref{fig:in-compatibility_side_by_side}(a), no dilatation (i.e., longitudinal) strain is generated. Interestingly, the energy cost encoded in  \eqref{eq:effective_nematic_free_energy_density} is even larger when the momentum is along the  the $[001]$ axis. In this case,  $\varphi_{2xy}$  completely overlaps with  $\Phi_4$, which in turn does not couple to the compatible strains. As shown in Fig. \ref{fig:in-compatibility_side_by_side}(c), the strain generated in this scenario deforms each volume element in a way that is discontinuous, such that cracks and overlaps develop along the boundaries of these elements.  We note that direction-selective criticality for the Ising-nematic order parameter was pointed out before \cite{Karahasanovic16,Paul17} and attributed to the anisotropic phonon dispersions of a crystal. Our results instead show that direction-selective nematic criticality is a universal behavior that arises from the compatible geometry of deformation and not from the anisotropy of the crystal, since in our case the underlying crystal is isotropic.

Since only $\Phi_{2,3}$ are critical in nemato-elasticity, one must elucidate the role of the non-critical $\Phi_{1,4,5}$ amplitudes which, as shown in Fig. \ref{fig:in-compatibility_side_by_side}(c), can generate incompatible strain patterns. The violation of the compatibility relations in continuum elasticity, which puts the system in the plastic regime, is due to crystalline defects \citep{eshelbyContinuumTheoryLattice1956, dewitLinearTheoryStatic1970, Dewit1973, dewitTheoryDisclinationsIII1973, dewitTheoryDisclinationsIV1973, kronerekkehartContinuumTheoryDefects1981, kleinertGaugeFieldsSolids1989, valsakumarGaugeTheoryDefects1988}. Mathematically, the latter are described in terms of a non-zero incompatibility tensor $\text{inc}(\varepsilon)_{ij}$, which can be readily incorporated in our helical formalism for nemato-elasticity (details in \citep{LongPaper}). 

\begin{figure}
    \centering
    \includegraphics[width=1.0\columnwidth]{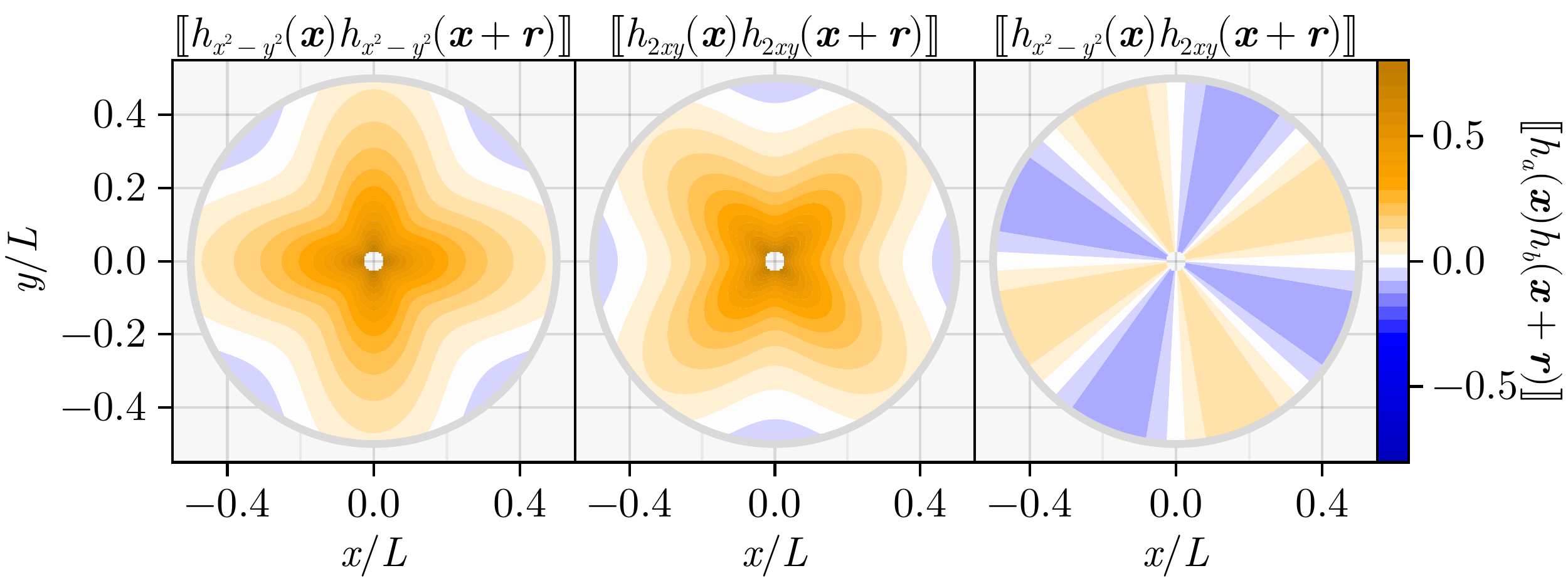}
    \caption{\justifying Normalized real-space strain-strain correlation functions calculated from an ensemble of quenched, independent, randomly distributed straight edge dislocations, in an isotropic medium of linear size $L$. These functions are computed explicitly in Ref. \citep{LongPaper}, where the double-bracket $\llbracket \cdot \rrbracket$ denotes the disorder-average. In order to satisfy the compatibility relations in disordered media, anisotropic spatial autocorrelations develop for the $h_{x^2 - y^2}$ and $h_{2xy}$ elastic strains (left and center panels, respectively). Additonally, radially independent anisotropic cross-correlations develop between $h_{x^2 -y^2}$ and $h_{2xy}$ despite their distinct symmetries (right panel), showing that both longitudinal and transverse random nematic fields emerge from crystal defects.}
    \label{fig:real-space-strain-correlations}
\end{figure}

The key point is that, in the presence of defects, the combination $\mathscr{E}_{ij} \equiv \varepsilon_{ij} - \varepsilon_{ij}^p$ is gauge-invariant, whereas the compatible strain $\varepsilon_{ij}$  and the incompatible ``plastic strain" $\varepsilon_{ij}^p$ are not. Since the free-energy must be gauge invariant, the nemato-elastic bilinear from Eq. \ref{eq:iso_nematoelastic_bilinear} is replaced by $ \boldsymbol{\varphi}\cdot \boldsymbol{\mathscr{E}} $. Minimization of the free energy with respect to the three compatible helical strain components yields the  \textit{nemato-plastic} contribution to the free energy \citep{LongPaper}:
\begin{equation}
    f_{\text{np}} = \tfrac{2\lambda_0}{\sqrt{3}}(1-\varrho) \Phi_1^{\phantom{p}}\Delta_p +\lambda_0(\Phi_4^{\phantom{p}}\varepsilon_4^p + \Phi_5^{\phantom{p}}\varepsilon_5^p), 
\end{equation}
where  $\varepsilon^p_{a}$  are the plastic strain components in the helical basis and $\Delta_p \equiv \frac{\sqrt{3}}{2}\varepsilon^p_1 - \varepsilon^p_0$ is the incompatible dilatation. Remarkably, this equation implies that strains generated by lattice defects couple only to the non-critical electronic nematic modes  $\Phi_{1,4,5}$. In contrast, nematic criticality is preserved in the compatible $\Phi_{2,3}$  nematic subspace, since these degrees of freedom are blind to crystalline defects. Thus, the minimization of  \eqref{eq:effective_nematic_free_energy_density} can be interpreted as minimizing incompatible electronic nematicity.

It is instructive to change back to the $d$-orbital basis to elucidate the impact of nemato-plasticity in terms of the nematic order parameter $\boldsymbol{\varphi}$. Performing this transformation, which is non-local in real space, reveals the emergence of the effective conjugate nematic field:
\begin{equation}
    \begin{aligned}
        \boldsymbol{h}_{\boldsymbol{q}} &\equiv -\left\{\boldsymbol{\Gamma}_0(\hat{q})\varepsilon_{0,\boldsymbol{q}}^p + \Gamma(\hat{q}) \cdot \boldsymbol{\varepsilon}^p_{\boldsymbol{q}}\right\},
    \end{aligned}
\end{equation}
where the projection operators $\boldsymbol{\Gamma}_0(\hat{q}) \equiv -\frac{2}{\sqrt{3}}(1-\varrho)\boldsymbol{\hat{Q}}_1$ and $\Gamma(\hat{q}) \equiv (1-\varrho)\boldsymbol{\hat{Q}}_1^{\phantom{\text{T}}}\boldsymbol{\hat{Q}}_1^{\text{T}} + \boldsymbol{\hat{Q}}_4^{\phantom{\text{T}}}\boldsymbol{\hat{Q}}_4^{\text{T}} + \boldsymbol{\hat{Q}}_5^{\phantom{\text{T}}}\boldsymbol{\hat{Q}}_5^{\text{T}}$ extract only the physical contribution from the plastic strain $\varepsilon^p_{ij}(\boldsymbol{x})$. As we show in detail in \citep{LongPaper}, it turns out that the Fourier transformed $\boldsymbol{h}(\boldsymbol{x})$  is nothing but the elastic strain emanating from defects in the absence of electronic nematicity. Thus, since the nemato-plastic free-energy can also be expressed $\boldsymbol{\varphi}(\boldsymbol{x})\cdot \boldsymbol{h}(\boldsymbol{x})$, it implies that \textit{all} \textit{$d$-orbital} nematic order parameters experience a highly nonlocal disorder field due to defects. Assuming the defects to be quenched at low temperatures, the defect strains are long-ranged, spatially correlated, random fields for the electronic nematic order parameter. A crucial consequence of this result is that, in an elastic medium, the inhomogeneous strain generated by lattice defects cannot be modeled as a simple uncorrelated distribution of strain values, like what is often assumed in, e.g., the random-field Ising-model \cite{Carlson2006, nieQuenchedDisorderVestigial2014, vojtaPhasesPhaseTransitions2013, vojtaDisorderQuantumManyBody2019}. As an example, we compute the in-plane components $h_{x^2-y^2}$ and $h_{2xy}$  generated by an ensemble of straight edge dislocations randomly placed within an isotropic medium (details in Ref. \citep{LongPaper}). The corresponding real-space correlation functions are shown in Fig. \ref{fig:real-space-strain-correlations}, illustrating the correlated, long-range nature of the conjugate field. Importantly, the cross-correlation between $h_{x^2-y^2}$ and $h_{2xy}$ is unavoidable and demonstrates that defects necessarily create both longitudinal and transverse random nematic fields. Since correlated random fields have an even greater tendency to break-up long-range order in domains (when compared to the uncorrelated case) \citep{weinribCriticalPhenomenaSystems1983, nattermannInstabilitiesIsingSystems1983, vojtaPhasesPhaseTransitions2013, vojtaDisorderQuantumManyBody2019}, we expect that the local nematic order parameter $\boldsymbol{\varphi}(\boldsymbol{x})$ will generally be inhomogeneous in the presence of defects. This happens even at nemato-elastic critically due to the momentum-direction-dependent mapping between the $d$-orbital basis and the helical basis.

To conclude, electronic nematicity is a common and robust phase of matter because elasticity both promotes and protects it. While the rotational symmetry breaking associated with nematicity is driven by electronic correlations, the lattice is not a passive bystander in nematic criticality. Gauge constraints from elastic compatibility restrict which nematic modes soften at the transition, enforcing direction-selective criticality by suppressing the incompatible ones that directly couple to microscopic plastic deformations. These results are relevant more broadly for quantum materials, given that electronic nematicity often arises as a vestigial order associated with other phases of matter such as spin and charge density-waves \cite{Fradkin2019,Fernandes2019}, and given its connection to quantum criticality and unconventional superconductivity \cite{Lederer2015,Lederer2017,Klein2018}.

\textit{Acknowledgments.} The authors would like to thank J. Freedberg, P. Littlewood, J. Vi\~nals, A. Rastogi, A. Chakraborty, and R. Aquino for many insightful discussions.

 \bibliography{bibliography_nematics.bib, additional_references.bib}
\end{document}